\documentclass[doublespacing]{elsart}
\usepackage{mathrsfs}

\usepackage{amssymb}

\usepackage{lineno}

\begin{document}

\begin{frontmatter}


 \title{From $k$-SAT to $k$-CSP: Two Generalized Algorithms\thanksref{label1}}
 \author{Liang Li},
 \author{Xin Li},
 \author{Tian Liu},  
    \ead{lt@pku.edu.cn}
    \address{Key Laboratory of High Confidence Software Technologies (Peking University),
    Ministry of Education, CHINA \\
    Institute of Software, School of Electronics Engineering and Computer Science,
      Peking University, Beijing, 100871, CHINA }
 \author{Ke Xu} 
    \ead{kexu@nlsde.buaa.edu.cn}
    \address{ National Lab of Software Development Environment,
     School of Computers,  Beihang University,  Beijing 100083,  China.}
 \thanks[label1]{Partially supported by the National 973 Program of China (Grant No. 2005CB321901).}









\end{frontmatter}


\section{Introduction}

Constraint satisfaction problems (CSPs) models many important
intractable $\mathcal{NP}$-hard problems such as propositional
satisfiability problem (SAT) \cite{D}.
Algorithms with non-trivial upper bounds
on running time for restricted SAT with bounded clause length $k$
($k$-SAT) can be classified into three styles: DPLL-like, PPSZ-like
and Local Search \cite{DHIV}, with local search algorithms having
already been generalized to CSP with bounded constraint arity $k$
($k$-CSP) \cite{S}. We generalize a DPLL-like algorithm in its
simplest form and a PPSZ-like algorithm \cite{PPZ} from $k$-SAT to
$k$-CSP. As far as we know, this is the first attempt to use
PPSZ-like strategy to solve $k$-CSP, and before little work has been
focused on the DPLL-like or PPSZ-like strategies for $k$-CSP.

For the DPLL-like deterministic $k$-CSP algorithm, a recurrent
inequality is tightly solved to get a non-trivial upper bound
$O^*((d-\frac{d-1}{d^k})^n)$ on running time, where $n$ is the
number of variables and $d$ is the domain size of variables in
input. For the PPSZ-like randomized $k$-CSP algorithm, the
Satisfiability Coding Lemma \cite{PPZ} is extended to non-Boolean
case to show that with probability approaching 1, a satisfying
assignment can be found in time $O^*((d\sqrt[k]{\frac{d-1}{d}})^n)$.
$O^*$ indicates that some polynomial factor in $n$ is ignored in
big-$O$ natation.

CSP generalizes SAT in two aspects: each variable can have more than
two available values, and each constraint can have more than one
falsifying partial assignments. These falsifying partial assignments
to some constraint are called \emph{nogoods}. For example, in graph
3-coloring problem a constraint of two variables $x$ and $y$ with
domain $\{0,1,2\}$ has tuples of values such as $(x:0,y:0)$,
$(x:1,y:1)$ and $(x:2,y:2)$ as its nogoods. If the first $k-1$
variables in a constraint with arity $k$ have values in agreement
with a nogood, then the last variable cannot have the value
specified by the nogood, so as not to falsify this constraint. Such
a variable is thus called \emph{narrowly chosen}. Our
generalizations rooted from the key observation that \emph{nogoods
(instead of constraints) in CSP can be treated as clauses in SAT to
produce narrowly chosen variables} which can be exploited by
algorithms to reduce their search efforts.

The remainder of the paper is organized as follows. Section 2
describes and analyzes the generalized DPLL-like deterministic
$k$-CSP algorithm. Section 3 extends the original satisfiability
coding lemma \cite{PPZ} to non-Boolean case. Section 4 presents the
PPSZ-like randomized $k$-CSP algorithm and its analysis.

\section{The DPLL-like Deterministic $k$-CSP Algorithm}

Our DPLL-like algorithm for $k$-CSP with variables domain size $d$
works as follow: for any nogood $(u_1:a_1,...,u_k:a_k)$, branch on
$u_1$ to $d-1$ branches, on each branch a value other than $a_1$ and
also different from value assigned on other sister branches is
assigned to $u_1$ and then recursively go down the branch. If all
these branch fails to find a satisfying assignment, then fix $u_1$
to value $a_1$ and branch on $u_2$ in exactly the same way as on
$u_1$ except that this time the number of remaining variables
decreased by one. Denote the running time of this algorithm by
$T(n)$, then clearly for $d \ge 2$ and $k \ge 2$:
$$T(n) \le (d-1)(T(n-1)+. . . +T(n-k))+\textrm{poly}(n).
\textrm{ }\textrm{ }\textrm{ }\textrm{ }\textrm{ }\textrm{ } (1) $$
Note that as usual we can safely ignore the additive poly$(n)$ term
at right hand side and treat the inequality as an equation.

When $d$ is a fixed constant, linear recursion (1) has solution
$T(n)=O^*(\lambda^n)$ with $\lambda$ the maximum root in
characteristic equation $f(\lambda)=\lambda^k-(d-1)(\lambda^{k-1}+.
. . +1)=0$. Since $\lambda>1$, our trick is to find the maximum root
in equation
$g(\lambda)=(\lambda-1)f(\lambda)=\lambda^{k+1}-d\lambda^k+(d-1)=0$.
$g(\lambda)$ is strictly increasing when
$\lambda>d(1-\frac{1}{k+1})$. We can find that when $\lambda \ge
d-\frac{d-1}{d^k}$, $g(\lambda)>0$; when
$\lambda=d-\frac{1}{d^{k-1}}$, $g(\lambda)<0$. Hence the tight
solution of (1) is $T(n)=O^*((d-\frac{d-1}{d^k})^n)$.

When $d$ is not fixed and varies with $n$, specifically $d=n^\alpha$
with $\alpha$ a constant, this case models some practical problems
(e.g. the Latin square problem and the $N$-queen problem) and a
random CSP model (called Model RB ), which contains many hard
instances seemingly quite challenging for various kinds of
algorithms, both theoretically \cite{XL} and experimentally
\cite{XBHL}, and a trivial upper bound is $O^*(n^{\alpha n})$.
Rewrite the recursion (1) as
$$ T(n) = (n^\alpha-1)(T(n-1)+...+T(n-k)).
\textrm{ }\textrm{ }\textrm{ }\textrm{ }\textrm{ }\textrm{ } (2) $$
When $\alpha \le 1$, for any fixed $\epsilon>0$, for large $n$ with
$n^\alpha-1>\frac{1}{\epsilon}$,  we have
$\sum_{i=n-k}^{n-1}T(i)<\epsilon T(n)$, so for large enough $n$
(actually $n+1$ will be fine for above $n$): $T(n) \le
(n^\alpha-1)(T(n-1)+\sum_{i=n-k}^{n-2}T(i))
<(n^\alpha-1)(T(n-1)+\sum_{i=n-k-1}^{n-2}T(i))
<(n^\alpha-1)(T(n-1)+\epsilon T(n-1))<n^\alpha(1+\epsilon)T(n-1)$.
Substitute $n$ by smaller numbers and combine these inequalities, we
have for any fix number $\epsilon>0$:
$T(n)=O^*((n!)^\alpha(1+\epsilon)^n)=O^*((\frac{n}{e})^{\alpha
n}(1+\epsilon)^n)$.

When $\alpha>1$, there is some number $\beta$ with $1<\beta<\alpha$,
such that for sufficiently large $n$, $(n-1)^\alpha-1>n^\beta$,  so
 $T(n) \le (n^\alpha-1)(T(n-1)+\sum_{i=n-k}^{n-2}T(i))
<(n^\alpha-1)(T(n-1)+\sum_{i=n-k-1}^{n-2}T(i))
=(n^\alpha-1)(T(n-1)+\frac{1}{(n-1)^\alpha-1}T(n-1))
<n^\alpha(1+\frac{1}{n^\beta})T(n-1) $. Since
$\prod_{n=1}^{\infty}(1+\frac{1}{n^\beta})$ converges to a finite
number,  by applying the same analysis as in above paragraph, we
have $T(n)=O^*((\frac{n}{e})^{\alpha n})$.

\section{A Generalized Satisfiability Coding Lemma}

Abbreviation \emph{w.r.t.} means \emph{with respect to}. Our key
generalization to a definition in \cite{PPZ} about \emph{isolated
points}, \emph{critical point} and \emph{critical variables} is:

\textbf{Definition 1}. For a $k$-CSP instance $F$ with domain $D$
for its $n$ variables, call $X=(a_1, ..., a_i, ..., a_n)$ an
\emph{isolated point} w.r.t. a set $S \subseteq D^n$ if there exists
a dimension $i \in \{1,2...,n\}$ and an $a'_i \in D-\{a_i\}$ such
that $X \in S$ but $X'=(a_1, ..., a'_i, ..., a_n) \not\in S$. Call
such a dimension $i$ a \emph{critical point} of $X$ w.r.t. $S$ and
the variable $u_i$ at dimension $i$ a \emph{critical variable}.

We only require that \emph{there exist} $a'_i \in D-\{a_i\}$ such
that $X \in S$ but $X'=(a_1, ..., a'_i, ..., a_n) \not\in S$, rather
than that \emph{for all} $a'_i \in D-\{a_i\}$ (which can only work
for SAT but not for CSP). This right choice (which works for both
SAT and CSP) makes the following two generalized lemmas and the
generalized algorithm with analysis in next section straightforward
to follow the routine in \cite{PPZ}, as follows.

Denote the number of critical points of $X$ w.r.t. $S$ by
$J_{S}(X)$. Call $X$ \emph{j-isolated} w.r.t. $S$ if $X$ is an
isolated point in exactly $j$ dimensions w.r.t. $S$. Call an
$n$-isolated solution $X$ an \emph{isolated solution}. When $S$ is
the set of all solutions of $F$, we can omit the words \emph{w.r.t.
$S$}. When solution $X=(a_1, ..., a_i, ..., a_n)$ has a critical
point $i$, there must be a constraint with a nogood in agreement
with $X$ except only in flipping $a_i$ to some $a'_i\in D-\{a_i\}$.
Call such a constraint \emph{critical}. In any value assigning
sequence of variables, if a critical variable $u_i$ is assigned
value last among all the variables in its critical constraint, and
all other variables than $u_i$ are assigned values in agreement with
$X$, then the value $a'_i$ should not be assigned to $u_i$
(otherwise the critical constraint will be falsified), thus the
domain of $u_i$ is narrowed. Call such a variable $u_{i}$
\emph{narrowly chosen}, otherwise \emph{fully chosen}. For any given
partial assignment and any constraint, we can efficiently check if a
variable in this constrain is narrowly chosen: it is narrowly chosen
iff other variables in this constrain has assigned values in
agreement with a nogood for this constraint, and every constraint
with arity $k$ can have at most $d^k$ nogoods.

\textbf{Lemma 1} Let $F$ be a $k$-CSP instance with a $j$-isolated
solution $X$. Then over all value assigning sequences of variables
with the final value assignment $X$, the average number of narrowly
chosen variables is at least $j/k$, thus the average number of fully
chosen ones is at most $n-j/k$.

\textbf{Proof:} (As in \cite{PPZ})
For a random value assigning sequence $\sigma$, since no constraint
involves more than $k$ variables in a $k$-CSP instance, the
probability that a critical variable is assigned last among all the
variables in its critical constraint is at least $1/k$. For each
critical constraint, if the corresponding critical variable is last
assigned, then this variable will be narrowly chosen. The
$j$-isolated solution $X$ has exactly $j$ critical points and these
$j$ critical variables each has a critical constraint. Thus, the
average number of narrowly chosen variables is at least $j/k$ when
$X$ is the final assignment. With a total number of variables $n$,
the average number of fully chosen variables is no more than
$n-j/k$. Q.E.D.

\textbf{Lemma 2} If a nonempty set $S\subseteq D^{n}$ with $|D|=d$,
then $\sum_{x\in S}(\frac{1}{d})^{n-J_{s}(x)} \geq 1$.

\textbf{Proof:} (By induction on $n$ as in \cite{PPZ}.)  Case $n =
0$ is trivially true. For $n>0$, consider a fixed dimension, say
$n$. Assume $D=\{a_1, ..., a_d\}$ and divide the set $S$ into $d$
subsets $S_1, ..., S_d$, such that $S_i = S'_i \times \{a_i\}$ with
$S'_i$ the projection of $S_i$ to the first $n-1$ dimensions. For
any $X$ in nonempty $S_i$, denote the image of $X$ in $S'_i$ by
$X'$, then induction hypothesis says $\sum_{x\in
S'_{i}}(\frac{1}{d})^{n-J_{S'_{i}}(x)-1} \geq 1$. Since $S$ is
nonempty, some $S_j$ is nonempty. For any $X \in S_j$, dimension $n$
is surely a critical point of $X$ w.r.t. $S_j$, so $J_{S_j}(X) =
J_{S'_j}(X') + 1$. On the other hand, dimension $n$ is a critical
point of $X$ w.r.t. $S$ iff some $S_i$ is empty. Say $S_i$ is empty,
then dimension $n$ is a critical point of $X$ w.r.t. $S$, so $J_S(X)
= J_{S_j}(X)$. In this case $\sum_{x\in S}(\frac{1}{d})^{n-J_S(x)}
\geq \sum_{x\in S_j}(\frac{1}{d})^{n-J_S(x)} = \sum_{x\in
S_j}(\frac{1}{d})^{n-J_{S_j}(x)} = \sum_{x\in
S'_j}(\frac{1}{d})^{n-J_{S'_j}(x)-1} \geq 1$. If no $S_{i}$ is
empty, then dimension $n$ is not a critical point of $X$ w.r.t. $S$,
so $J_S(X) = J_{S_i}(X)-1$. In this case $\sum_{x\in
S}(\frac{1}{d})^{n-J_S(x)} = \sum_{i=1}^{d}\sum_{x\in
S_i}(\frac{1}{d})^{n-J_S(x)} = \sum_{i=1}^{d}\sum_{x\in
S_i}(\frac{1}{d})^{n-J_{S_i}(x)+1 } = \sum_{i=1}^{d}\sum_{x\in
S'_i}(\frac{1}{d})^{n-J_{S'_i}(x)} =
\frac{1}{d}\sum_{i=1}^{d}\sum_{x\in
S'_i}(\frac{1}{d})^{n-J_{S'_i}(x)-1} \geq \frac{1}{d}\sum_{i=1}^{d}1
= 1$. Q.E.D.

\section{ PPSZ-like Randomized $k$-CSP Algorithm }

Our PPSZ-like algorithm for $k$-CSP and its analysis generalize from
one for $k$-SAT \cite{PPZ} with the key observation that we can use
a partial assignment and nogoods to efficiently produce narrowly
chosen variables w.r.t. some value assigning sequence of variables,
as explained in introduction and last sections.

\newcommand{\tab}{\phantom{o}\hspace{4ex}}
\textbf{Algorithm A}\\
\textbf{repeat}  $n(n+1)(d\sqrt[k]{(d-1)/d})^n$  times\\
$\tab$\textbf{while} there exists an unassigned variable\\
$\tab$$\tab$select an unassigned variable $y$ at random\\
$\tab$$\tab$\textbf{if} $y$ is narrowly chosen \\
$\tab$$\tab$$\tab$\textbf{then} set $y$ to a random value in the narrowed domain\\
$\tab$$\tab$$\tab$\textbf{else} set $y$ to a random value in its full domain\\
$\tab$\textbf{if} the CSP instance is satisfied,  \textbf{then}
output the assignment

Now we prove that Algorithm A can find a solution to a satisfiable
$k$-CSP instance $F$ in time $O^*((d\sqrt[k]{\frac{d-1}{d}})^n)$
with probability approaching $1$. Suppose that $X$ is an
$j$-isolated solution of $F$ with $j$ critical points ($1\leq j \leq
n$, since $j=0$ is a trivial case of tautology input without any
nogood). In one iteration of the \textbf{repeat} loop, by lemma 1,
the average number of critical variables assigned last among all the
variables in its critical constraint is at least $j/k$, over the
random value assigning sequences of variables in the \textbf{while}
loop. Then by Markov inequality (on complement event), the
probability of the event that for at least $j/k$ critical
constraints, the critical variables occur last among the variables
in the critical constraint, is at least $\frac{1}{n - j/k + 1}$.
When this event occurs, the number of fully (narrowly) chosen
variables is at most $n-j/k$ (at least $j/k$), and each fully
(narrowly) chosen variable's value has probability exact
$\frac{1}{d}$ (at least $\frac{1}{d-1}$) to agree with the
corresponding value of $X$, so the probability of the event that the
values assigned to the variables in \textbf{while} loop agree with
the assignment $X$ is at least
$(\frac{1}{d})^{n-j/k}(\frac{1}{d-1})^{j/k}$ conditioned on the
above event. Thus, the probability that a $j$-isolated solution $X$
of $F$ is output by algorithm A is at least $\frac{1}{n- j/k
+1}(\frac{1}{d})^{n-j/k}(\frac{1}{d-1})^{j/k}$. By summing up this
probability over set $S$ of all solutions of $F$ and by lemma 2, the
probability that algorithm A outputs some solution is at least
$\sum_{x\in
S}\frac{1}{n-J_{S}(X)/k+1}(\frac{1}{d})^{n-J_S(x)/k}(\frac{1}{d-1})^{J_{S}(X)/k}
\geq$ $ \frac{1}{n+1}(\frac{1}{d})^{n-n/k} (\sum_{x\in
S}(\frac{1}{d})^{n-J_{S}(X)})^{1/k}(\frac{1}{d-1})^{J_{S}(X)/k} \geq
\frac{1}{n+1}(\frac{1}{d})^{n-n/k}\cdot 1 \cdot
(\frac{1}{d-1})^{n/k} =
\frac{1}{n+1}(d\sqrt[k]{\frac{d-1}{d}})^{-n}$. So by repeating the
\textbf{while} loop $n(n+1)(d\sqrt[k]{\frac{d-1}{d}})^n =
O^*((d\sqrt[k]{\frac{d-1}{d}})^n)$ times, we can find a satisfying
assignment with probability approaching 1. When $d=n^{\alpha}$, this
upper bound becomes $O^*(n^{\alpha n (1- \frac{1}{kn^{\alpha}\ln
n})})$.


\section{Conclusion and Future Work}

We have generalized two algorithms from $k$-SAT to $k$-CSP, with
running time better than the trivial bound $O^*(d^n)$ when variable
domain size $d$ is fixed. When $d$ is unfixed, say $d=n^\alpha$, the
result is only slightly better than the trivial bound $O^*(n^{\alpha
n}$), whether we can reach $O^*(n^{\beta n}$) (where $\beta<\alpha$
is a constant) in this case is still open. Our solutions to the
recursion (1) and (2) might find other application in the analysis
of DPLL-like algorithms. Our randomized algorithm is the first
application of PPSZ-like strategy beyond SAT to CSP.

In summary, this paper can be viewed as the first step toward
establishing upper bounds for solving $k$-CSP using DPLL-like or
PPSZ-like strategies, which leaves mcuh room for further study and
improvement, for example, by combining PPSZ-like and local search
algorithms as in \cite{IT}.



\end{document}